\documentclass[aps,prb,twocolumn,floatfix,groupedaddress,showpacs]{revtex4}

\newcommand{\nvp}{$n_{v}$ }
\newcommand{\ab}{\emph{ab initio }}

\newcommand{\etal}{\emph{et al.} }

\usepackage{placeins}
\usepackage{latexsym}
\usepackage{bm}
\usepackage{graphicx,epsf}
\usepackage{dcolumn}
\usepackage{times}
\usepackage{psfrag}
\usepackage{color}

\begin{document}

%\preprint{}

\title{Band structures of rare gas solids within the $GW$ approximation}

\author{S. Galami\'{c}-Mulaomerovi\'{c}}
\author{C.H. Patterson}

\affiliation{ Department of Physics and Centre for Scientific Computation,\\
University of Dublin, Trinity College, Dublin 2, Ireland}

\date{\today}

\begin{abstract}
Band structures for solid rare gases (Ne, Ar) have been calculated using the $GW$ approximation. All electron and pseudopotential \ab calculations were performed using Gaussian orbital basis sets and the dependence of particle-hole gaps and electron affinities on basis set and treatment of core electrons is investigated. All electron $GW$ calculations have a smaller particle-hole gap than pseudopotential $GW$ calculations by up to 0.2 eV. Quasiparticle electron and hole excitation energies, valence band widths and electron affinities are generally in very good agreement with those derived from optical absorption and photoemission measurements.
\end{abstract}



\maketitle


\section{\label{Introduction:S}Introduction}

Optical spectra and band structures of rare gas solids (RGS) have been studied, both experimentally and theoretically, for over 40 years. Their importance lies in the simplicity of their crystal structure, the nearly atomic character of valence states versus extended character of conduction states and the fact 
that they have strong many-body effects in their optical spectra. They are an important testing ground for electronic structure methods and as electronic structure methods have developed, they have been applied to RGS. Early electronic structure studies included applications of density functional theory 
(DFT) \cite{Lipari70,Bacalis88,Magyar04}, Hartree-Fock theory (HFT) \cite{Dagens72,Kunz73,Baroni81,Baroni84} and self-interaction corrected DFT \cite{Heaton83}. Some of these studies have included correlation effects 
in the band structure via many-body perturbation theory \cite{Lipari70,Kunz73,Baroni84,Bacalis88,Rohlfing00}. 

In this paper we present results of \ab all electron and pseudopotential many-body calculations of the band structures of solid Ne and Ar using Gaussian orbital basis sets. Band structures are calculated using the $GW$ approximation \cite{Hedin69,Hybertsen86} and the dependence of particle-hole gaps and electron affinities on basis set and treatment of core electrons is investigated. In another paper \cite{Galamic04a} we will present results of calculations of optical spectra of these solids using a Bethe-Salpeter formalism \cite{Rohlfing00}. Calculations were performed in a Gaussian orbital basis using the EXCITON code \cite{Galamic04}, which is interfaced to the CRYSTAL code \cite{Crystal03}. Single particle wave functions, energy eigenvalues and matrix elements of the exchange correlation potential from CRYSTAL are used by EXCITON to perform $GW$ and exciton calculations. The principal parameters of the band structures of RGS which have been obtained experimentally are the particle-hole band gap, $E_G$, the valence band width, $W_V$ and the electron affinity, $E_A$. 
The particle-hole gap is the energy difference for particle and hole excitations at the conduction band minimum and valence band maximum. 
It has been obtained experimentally from absorption spectrum \cite{Runne95} and photoemission measurements \cite{Schwentner73,Schwentner75}. 

The $GW$ approximation is a many-body perturbation theory and therefore contains corrections to a simpler, single-particle (SP) Hamiltonian. It was originally applied to semiconductors using a DFT Kohn-Sham Hamiltonian, a plane wave basis set and pseudopotential (PP) approximation for core electrons \cite{Hybertsen86} and was found to give excellent agreement with experiment for relatively narrow band gap materials such as Si, where DFT results in an indirect band gap which underestimates the experimental value by $\sim 0.7$~eV. 

All electron $GW$ calculations have been performed recently for a variety of crystalline solids, including Si \cite{Hamada90,Arnaud00,Kotani02,Ku02}. The all electron indirect band gaps calculated for Si were underestimated \cite{Hamada90,Kotani02,Ku02} by 0.2 to 0.3 eV. There has been some debate whether this is due to incompleteness of the basis \cite{Tiago04} or explicit inclusion of the core electrons (all electron rather than PP approximation) \cite{Hamada90,Arnaud00,Kotani02,Ku02,Lebegue03}. In the present work we compare results for the RGS using both all electron and PP approximations for the core electrons.

The remainder of this paper is organized as follows: in section II the $GW$ formalism used here is outlined, in section III results of $GW$ bandstructure calculations are compared to experiment and earlier $GW$ calculations on Ne and Ar. Finally, conclusions are given in section IV.

\section{\label{Theory:S}GW Approximation}

\subsection{Quasiparticle energies}

Conceptually, single electron and hole excitations are described %within the $GW$ approximation as electrons or holes scattering from plasmons to form quasiparticles. These virtual scattering events shift quasiparticle energies (QPE) relative to their DFT values and modify SP wave functions to yield quasiparticle amplitudes. The Schr\"odinger-like equation which determines quasiparticle amplitudes, $\psi _{m} ^{QP}(\bm{r})$ and energies, $\epsilon _{m}$, of the state $m$ is \cite{Hedin69}
\begin{equation}
\label{eqn1}
H(\bm r)\psi _{m} ^{QP}(\bm{r}) + \int \Sigma (\bm{r},\bm{r'},E)\psi _{m} ^{QP}(\bm{r'})\mathrm{d}\bm{r'} = \epsilon _{m} \psi _{m} ^{QP}(\bm{r}).
\end{equation}
The self-energy operator, $\Sigma (\bm{r},\bm{r'},E)$, is non-Hermitian and so eigenvalues, $\epsilon_{m}$, have real and imaginary parts, the real part being the quasiparticle energy, $E^{QP} _{m}$ and the imaginary part being related 
to the quasiparticle lifetime. In this work the self-energy operator was computed within the $GW$ approximation \cite{Hedin65}, in which the 
self-energy operator is obtained from convolution of the non-interacting single-particle Green's function, $G_o$, and the screened Coulomb interaction, $W$,
\begin{equation}
\label{eqn2}
\Sigma (\bm{r}, \bm{r'},E) = \frac{\imath}{2\pi } \int \mathrm e^{-\imath\omega 0^{+}} \, G_{o}(\bm{r},\bm{r'},E-\omega )\,
W(\bm{r},\bm{r'},\omega )\,\mathrm{d} \omega.  
\end{equation}
$G_{o}$ is constructed from DFT single-particle orbitals, $\psi_{n} ^{SP}$, and eigenvalues, $E_{n} ^{SP}$, of the Kohn-Sham operator, 
\begin{equation}
\label{eqn3}
G_{o}(\bm{r}, \bm{r'}, \omega ) = \sum_{n} \frac{\psi_{n} ^{SP}(\bm{r}) \psi_{n} ^{\ast SP}(\bm{r'})}{\omega -E_{n} ^{SP}+\imath 0^{+}\mathrm{sign}(E _{n} ^{SP}-E_{F})}.
\end{equation}

For light elements it has been found that quasiparticle amplitudes are well approximated by DFT wave functions \cite{Hybertsen86}. Thus, quasiparticle energies are simply given by
\begin{equation}
\label{energyqp:E}
E^{QP}_{m\bm{k}} = E^{SP}_{m\bm{k}} + \langle \psi_{m\bm{k}}^{SP} | \Sigma(E^{QP}_{m\bm{k}}) -V_{xc}[n_{v}]| \psi_{m\bm{k}} ^{SP} \rangle.
\end{equation}
In this case only diagonal elements of the self-energy matrix and exchange-correlation potential, $V_{xc}[n _{v}]$, are required. $V_{xc}[n_{v}]$ is the exchange-correlation potential of the valence electron density, $n_{v}$, from
the initial DFT calculation. Equation~(\ref{energyqp:E}) is solved using a scheme given by Hybertsen and Louie \cite{Hybertsen86}.
\begin{equation}
\label{eqn5}
E^{QP}_{m\bm{k}} \approx E^{SP}_{m\bm{k}} + Z_{m\bm{k}}\Re \langle \psi_{m\bm{k}}^{SP} | \Delta\Sigma(E^{SP} _{m\bm{k}})| \psi_{m\bm{k}} ^{SP} \rangle.
\end{equation}
$Z_{m\bm{k}}$ is the quasiparticle renormalization factor and is defined by
\begin{equation}
\label{eqn6}
Z_{m\bm{k}} = \left( 1 - \frac{\partial \Re \langle \psi_{m\bm{k}}^{SP} | \Sigma_c(E) | \psi_{m\bm{k}} ^{SP} \rangle }{\partial E} \big | _ {E = E^{SP}_{m\bm{k}}}  \right)^{-1} 
\end{equation}
and the operator $\Delta\Sigma$ is given by
\begin{equation}
\label{eqn7}
\Delta\Sigma(\bm{r},\bm{r'},E) = \Sigma(\bm{r},\bm{r'},E) - V_{xc}[n_{v}].
\end{equation}
\subsection{Self-energy matrix elements}

The screened Coulomb interaction is computed from the dielectric function and the bare Coulomb interaction
\begin{equation}
\label{eqn9}
W(\bm{r},\bm{r'}, \omega) = \int \epsilon^{-1} (\bm{r},\bm{r''}, \omega)  v(\bm{r''},\bm{r'}) \mathrm{d}\bm{r''}
\end{equation}
Two-point functions in a crystal lattice such as the screened interaction have the property, $f(\bm{r}+\bm{R}, \bm{r'}+\bm{R}) = f(\bm{r}, \bm{r'})$, where $\bm{R}$ is a Bravais lattice vector, owing to translational symmetry. They may be represented as a Fourier transform as \cite{Hybertsen86},
\begin{equation}
\label{eqn10}
f(\bm{r}, \bm{r'}) = \sum _{\bm{q}, \bm{G}, \bm{G'}} 
e^{\imath (\bm q+\bm G)\cdot \bm r} f_{\bm G \bm{G'}}(\bm q) e^{-\imath (\bm q+\bm G')\cdot \bm r'},
\end{equation}
where $\bm G$ is a~reciprocal lattice vector and $\bm q$ is a wavevector in the first Brillouin zone. Fourier coefficients of the screened potential, $W_{\bm G\bm G'}$ are given by,
\begin{equation}
\label{eqn11}
W_{\bm G\bm{G'}}(\bm q,\omega) = \frac{4\pi e^{2}}{\Omega} \frac{1}{|\bm q+\bm G||\bm q+\bm{G'}|} \varepsilon ^{-1}_{\bm G\bm{G'}}(\bm q, \omega), 
\end{equation}
where $\Omega $ is the crystal volume and $\varepsilon _{\bm G\bm{G'}}^{-1}$ is the inverted, symmetrized RPA dielectric matrix. Numerical evaluation of W and $\Sigma $ (Eqs.~(\ref{eqn2}) and~(\ref{eqn11})) requires calculation and inversion of the dielectric matrix at many values of $\omega$. Although such schemes have been carried out \cite{Godby88}, it is both time consuming and unnecessary in the present work. Instead, we adopt a plasmon-pole model based on the work of von der Linden and Horsch \cite{vonderLinden88}, which uses the concept of dielectric band structure \cite{Baldereschi79} to approximate the frequency dependence of the dielectric matrix. The model assumes that all frequency dependence is projected onto eigenvalues of the inverted 
dielectric matrix, $\varepsilon ^{-1} _{\bm q l}(\omega)$ through the approximation,
\begin{equation}
\label{eqn12}
\varepsilon ^{-1} _{\bm q l}(\omega) = 1+ \frac{z_{\bm ql}\omega _{\bm ql}}{2}
\Bigg(\frac{1}{\omega - \omega _{\bm ql} + \imath 0^{+}}-\frac{1}{\omega + \omega _{\bm ql} - \imath 0^{+}}\Bigg).
\end{equation} 
$z_{\bm ql}$ are pole strengths, $\omega _{\bm ql}$ are plasmon frequencies and $0^{+}$ is a positive infinitesimal. Eigenvalues of the inverted dielectric matrix determine the pole strengths and a plot of their dispersion with wave vector is known as the dielectric band structure \cite{Baldereschi79}; the dielectric band structure for fcc Ar was reported previously \cite{Galamic01}. Plasmon pole frequencies are calculated either using the Johnson sum 
rule \cite{Johnson74} or by fitting the dielectric matrix at zero frequency 
and a finite, imaginary frequency, $\imath \omega _{f}$. It can easily be shown that,
\begin{equation}
\label{eqn13}
\omega ^{2} _{\bm ql}= \frac{-\omega ^{2} _{f}[1-\varepsilon ^{-1} _{\bm{q}l}(\imath \omega _{f})]}
{\varepsilon ^{-1} _{\bm{q}l}(\imath 0)-\varepsilon ^{-1} _{\bm{q}l}(\imath \omega _{f})},
\end{equation}  
and that, 
\begin{equation}
\label{eqn14}
z_{\bm ql}=1-\varepsilon ^{-1} _{\bm{q}l}(\imath 0).
\end{equation}

Results presented in Section~\ref{results:S} were obtained using plasmon frequencies from the Johnson sum rule; 
quasiparticle band gaps obtained in this way were generally higher than those obtained by fitting by a few 
hundredths of an electron Volt. The plasmon-pole form of the inverted dielectric matrix allows the 
frequency integration in the calculation of the self-energy matrix element in Eq.~(\ref{eqn2}) to be done 
analytically. This leads to two contributions to the self-energy: an energy independent, Hartree-Fock exchange term, 
\begin{eqnarray}
\label{eqn15}
\lefteqn{ \langle m\bm k|\Sigma _{x}|m \bm k\rangle  = } \nonumber \\
& & {\displaystyle -\frac{4 \pi ^{2}e^{2}}{\Omega } 
 \sum_{\bm q,\bm G}\sum_{n}^{occ} \frac{|\langle m \bm k|\mathrm e^{-\imath(\bm q+\bm G)\cdot \bm r}| n \bm k+\bm q \rangle | ^{2}}{|\bm q+\bm G|^{2}} },
\end{eqnarray}
where sum over bands, $n$, extends only over occupied states. The second, dynamic part
\begin{widetext}
\begin{eqnarray}
\label{eqn16}
 \langle \, m \bm k  | \Sigma_{c} (E)  | m\bm k \, \rangle = \frac{4 \pi ^{2}e^{2}}{\Omega } \sum _{\bm{q,G,G^{'}}} \sum _{n}^{occ} \frac{\langle m \bm k| \mathrm e^{-i(\bm q+\bm G)\cdot \bm r}| n \bm k+\bm q \rangle \, \langle n \bm k+\bm q | \mathrm e^{i(\bm q+\bm G^{'})\cdot \bm r'}|m \bm k \rangle }{|\bm q+\bm G|\, |\bm q+\bm G^{'}|}  \nonumber \\ 
\times \sum _{l}V_{l,-\bm{G}}(-\bm{q})V^{*} _{l,-\bm{G}^{'}}(-\bm{q})\bigg[ \frac{z_{-\bm{q}l}\omega _{-\bm{q}l}}{2}\frac{1}{E-E_{n\bm{k}+\bm{q}}+\omega _{-\bm{q}l}\mathrm{sign}(E_{F} - E_{n\bm{k}})}\bigg], 
\end{eqnarray}
\end{widetext}
contains correlation energies of electron or hole quasiparticles \cite{Hott91}. $V_{l,-\bm{G}}$ are eigenvectors 
of the static, symmetrized dielectric matrix $\varepsilon _{\bm G\bm G'}(\bm q, \omega =0)$. 

\subsection{Numerical details}

The starting point in our approach is to generate non-interacting single-particle Green's functions of an N-electron system.  We use density functional theory~\cite{KohnPR:136} (DFT) within the Perdew-Wang generalized gradient approximation~\cite{Perdew92} (PWGGA) to obtain eigenvectors and eigenvalues of the Kohn-Sham Hamiltonian. For this part of the calculation we employ the \emph{ab initio} package CRYSTAL \cite{Crystal03} which uses the Linear Combination of Atomic Orbitals (LCAO) approach to expand the Bloch functions. In order to investigate convergence criteria within the Gaussian orbital framework and effects of core electrons we performed both all electron and PP calculations for each solid. Basis sets with 52 and 53 functions per atom were developed for PP and all electron calculations, respectively, for Ne. The PP and all electron basis sets for Ar contained 56 and 66 functions, respectively. Details of basis sets are given in Appendix~\ref{basisset:A}. Pseudopotentials from Durand and Barthelat \cite{Barthelat77} were used in PP calculations.  Experimental lattice constants \cite{Sonntag77} were used. 

The sum over $\bm q$ points in Eqs.~(\ref{eqn15}) and  (\ref{eqn16}) as well as integration over the Brillouin zone in the dielectric matrix calculation is performed using Monkhorst-Pack \cite{Monkhorst76} special points. The singularity in Eqs.~(\ref{eqn15}) and (\ref{eqn16}) of  $1/q^{2}$ type for  $\bm q\to 0$ and $\bm G=\bm G'=0$ was integrated out using the auxiliary function technique of Gygi and Baldereschi \cite{Gygi86}, while the singularity in Eq.~(\ref{eqn16}) of $1/q$ type was neglected since the final result is not affected if it is neglected \cite{Arnaud00}.

Two special points in the irreducible Brillouin zone were used for calculation of self-energy matrix elements and an $8\times 8\times 8$ grid in the full Brillouin zone was used for the dielectric matrix calculation. Up to 400 (8000) $\bm G$ vectors are required to achieve convergence of the Hartree-Fock part of the self-energy (Eq.~(\ref{eqn15})) for PP (all electron) basis sets for second row elements. In the summation over $\bm G$ and $\bm{G'}$ vectors in Eq.~(\ref{eqn16}), 65 vectors gave well converged results for all solids. 

\subsection{Core-valence exchange-correlation decoupling}

When matrix elements of the $\Delta \Sigma$ operator in Eq. (\ref{eqn7}) are evaluated, contributions from core electrons to the valence electron self-energy must be considered \cite{Hedin69,Hybertsen86}. We compare results from two alternative approximations for the energy independent part of the $\Delta \Sigma$ operator which were applied recently in all electron GW calculations on Si \cite{Arnaud00}. The first approximation is to compute matrix elements of the DFT exchange-correlation potential using the valence electron density only and to restrict the sum on occupied states in Eq. (\ref{eqn15}) to valence states only,
\begin{eqnarray}
\langle m\bm k|\Delta \Sigma|m \bm k\rangle  = 
\langle m\bm k|\Sigma_{c}^{val}|m \bm k\rangle + 
\langle m\bm k|\Sigma_{x}^{val}|m \bm k\rangle \nonumber \\ - 
\langle m\bm k|V_{xc}[n_{v}]|m \bm k\rangle.
\label{Vxc_val:E}
\end{eqnarray} 
The second approximation is to replace matrix elements of the valence-density-only exchange correlation potential, $V_{xc}[n_{v}]$, in Eqs. (\ref{eqn7}) and (\ref{Vxc_val:E}) by
\begin{eqnarray}
\langle m\bm k|V_{xc}[n_{c}+n_{v}]|m \bm k\rangle
& - &
\langle m\bm k|\Sigma_{x}^{core}|m \bm k\rangle.
\label{Vxc_core:E}
\end{eqnarray} 
The notation $\langle m\bm k|\Sigma_{x}^{core}|m \bm k\rangle$ and $\langle m\bm k|\Sigma_{x}^{val}|m \bm k\rangle$ indicates that the sum on $n$ in Eq. (\ref{eqn7}) is limited to core or valence states only. Matrix elements of the LDA exchange potential and Hartree-Fock exchange operator for valence band maximum and conduction band minimum states in silicon obtained by Arnaud and Alouani \cite{Arnaud00} using a projector augmented wave (PAW) method and in this work using CRYSTAL are compared in Table \ref{Exchange:T}. Remarkably good agreement was found between LDA exchange potential matrix elements from either method (differences in matrix elements are only 0.01 eV in three out of four cases) and good agreement between Hartree-Fock exchange matrix elements is also obtained (within 0.1 eV). The shortcoming of the latter approach (Eq. (\ref{Vxc_core:E})) is slow convergence of the Hartree-Fock exchange operator for core states, $\langle m\bm k|\Sigma _{x}^{core}|m \bm k\rangle$. 

\begin{table}
\caption{\label{Exchange:T} Matrix elements of the Hartree-Fock exchange ($\Sigma_{x}$) and LDA exchange potential (V$_{x}$) operators for self-consistent DFT wave functions at valence band maxima and conduction band minima for Si. The symbols $n_{c}+n_{v}$, $core$ and \nvp denote whether core + valence, core only or valence only states are included in the operators.}
\begin{ruledtabular}
\begin{tabular}{c d d d}
      & \multicolumn{1}{c}{$V_{x}[n_{c}+n_{v}]$}
      & \multicolumn{1}{c}{$\Sigma_{x}^{core}$}
      & \multicolumn{1}{c}{$V_{x}[n_{v}]$}  \\
\hline
 $\Gamma_{15c}$\footnotemark[1]    & -11.75 & -1.32 & -10.18 \\
 $\Gamma_{15c}$\footnotemark[2]    & -11.74 & -1.40 & -10.19 \\
\\
$\Gamma_{25'v}$\footnotemark[1]     & -13.55 & -1.80 & -11.45 \\
$\Gamma_{25'v}$\footnotemark[2]    & -13.45 & -1.85 & -11.46 \\
\end{tabular}
\end{ruledtabular}
\footnotetext[1]{This work.}
\footnotetext[2]{Reference \onlinecite{Arnaud00}.}
\end{table}
\section{\label{results:S} Results and discussion}

One of the aims of this work is to compare results of $GW$ calculations on simple atomic solids which treat core electrons either by a PP or by explicitly including them in an all electron calculation.  Energies of states at valence band maxima and conduction band minima are given in Table~\ref{QPenergies:T}, as well as fundamental band gaps and valence band widths within DFT and $GW$ approximations and experiment. $GW$ all electron quasiparticle energies obtained by the two core-valence electron decoupling methods outlined in Eqs. (\ref{Vxc_val:E}) and (\ref{Vxc_core:E}) are given in columns labelled $E^{QP}_{(1)} $ and $E^{QP}_{(2)}$. DFT-PP calculations underestimate experimental band gaps by 45\% for Ne and by 35\% for Ar. $GW$-PP calculations show significantly improved agreement with experimental data in each solid; the band gap error is reduced to 6\% in Ne and 2\% in Ar; slight underestimation of band gaps in RGS is similar to that in semiconductors where, for example, the band gap is underestimated by 4\% in Si \cite{Shirley97}. The reason for good agreement between quasiparticle energies, $E^{QP}_{1}$ and $E^{QP}_{2}$, for Ne and Ar (this work) and Si \cite{Arnaud00} is that the energy independent part of the $\Delta \Sigma$ operator is $\Sigma_{x}^{val}$ - $V_{xc}[n_{v}]$ for $E^{QP}_{1}$ and it is 
$\Sigma_{x}^{core+val}$ - $V_{xc}[n_{c}+n_{v}]$ for $E^{QP}_{2}$; the difference in these two quantities is of order 0.1 eV and results in nearly equal quasiparticle energies, $E^{QP}_{1}$ and $E^{QP}_{2}$.
\begin{table}
\caption{\label{QPenergies:T}DFT eigenvalues and $GW$ quasiparticle energies at valence band maxima and conduction band minima, valence band widths and energy gaps for Ne and Ar RGS. Calculations were performed using pseudopotentials (second and third columns) and all electron basis sets (fourth to sixth columns). The fifth column, $E^{QP}_{(1)} $ gives all electron $GW$ data when $V_{xc}[n_{v}]$ is calculated explicitly (Eq.~(\ref{Vxc_val:E})) and the sixth column gives quasiparticle energies when $V_{xc}[n_{v}]$ is calculated using Eq.~(\ref{Vxc_core:E}). The last column gives experimental data. Experimental data is taken from Reference~\onlinecite{Runne95} unless cited differently. Energies are given in eV.}
\begin{ruledtabular}
\begin{tabular}{c d d d d d d}
 & 
 \multicolumn{2}{c}{PP} &%
 \multicolumn{3}{c}{all-electron} \\
&\multicolumn{1}{c}{DFT} &%
 \multicolumn{1}{c}{$E^{QP}$} &%
 \multicolumn{1}{c}{DFT} &%
 \multicolumn{1}{c}{$E^{QP}_{(1)}$} &%
 \multicolumn{1}{c}{$E^{QP}_{(2)}$} &%
 \multicolumn{1}{c}{Exp.} \\
\hline
\multicolumn{7}{l}{Neon} \\
$\Gamma _{15v}$ &  -13.14  &  -19.37  & -13.18  & -19.07  & -19.10  & -20.21  \\
$\Gamma _{1c}$  &   -1.35  &    0.86  &  -1.42  &   0.97  &   1.03  &   1.3   \\
$W_{v}$         &    0.71  &    0.93  &   0.79  &   0.91  &   0.93  &   1.3\footnotemark[1] \\
$E_{g}$         &   11.99  &   20.23  &  11.76  &  20.04  &  20.13  &  21.51  \\
\\
\multicolumn{7}{l}{Argon}\\
$\Gamma _{15v}$ &  -9.74 &   -13.15 &  -10.27 &  -13.02  & -13.00  &  -13.75 \\
$\Gamma _{1c}$  &  -0.60 &   0.72   &  -0.76  &    0.80  &   0.81  &  0.4  \\
$W_{v}$         &   1.35 &   1.73   &   1.32  &    1.83  &   1.85  &  1.7\footnotemark[1]\\
$E_{g}$         &  9.13  &   13.89  &   9.51  &  13.82   &   13.81 &  14.15 \\
\end{tabular}
\end{ruledtabular}
\footnotetext[1]{Reference \onlinecite{Schwentner75}.}
\end{table}

There is good agreement between PP and all electron DFT calculations for Ne; the bottom conduction ($\Gamma _{15v}$) and top valence band ($\Gamma _{1c}$) energies and valence band widths ($W_{v}$) differ by less than 0.1~eV. $\Gamma_{15v}$ and $\Gamma_{1c}$ quasiparticle energies from all electron calculations lie slightly above PP values. The absolute value of the $\Gamma _{1c}$ conduction band energy, which determines the electron affinity, has the correct sign in GW calculations and lies just 0.4 eV below the experimental value for the electron affinity, whereas DFT calculations predict larger electron affinities of the wrong sign. $GW$ calculations result in valence band widths $\sim$ 0.9 eV, which are smaller than the experimental value of 1.3 eV \cite{Schwentner75}, but are in agreement with the value of 0.99~eV obtained by Bacalis \etal \cite{Bacalis88}  The two methods used for core-valence decoupling (Table~\ref{QPenergies:T}, columns 5 and 6) result in $\Gamma _{15v}$  and $\Gamma_{1c}$ quasiparticle energies which differ by only $\sim$ 0.05eV. 
 
The $\Gamma _{15v}$ valence band maximum state in all electron DFT calculations on Ar is lower than in PP calculations by 0.53~eV while the $\Gamma _{1c}$ conduction band minimum state is lower by 0.16 eV. However $GW$ quasiparticle energies for these states using either all electron or PP basis sets are in good agreement, the maximum difference being only 0.13 eV. The $GW$ $\Gamma _{1c}$ conduction band energy exceeds the experimental electron affinity by $\sim$ 0.4 eV, whereas the DFT $\Gamma _{1c}$ energy again predicts an electron affinity with the wrong sign. $GW$ valence band widths of 1.73 (PP) and 1.83~eV (all electron) agree very well with the experimental value of 1.7~eV \cite{Schwentner75}. The two methods used for core-valence decoupling also result in very similar quasiparticle energies for Ar. 

\begin{figure*}
\psfrag{Ne}[cc][]{\colorbox{white}{Ne}}
\includegraphics[height=8 cm]{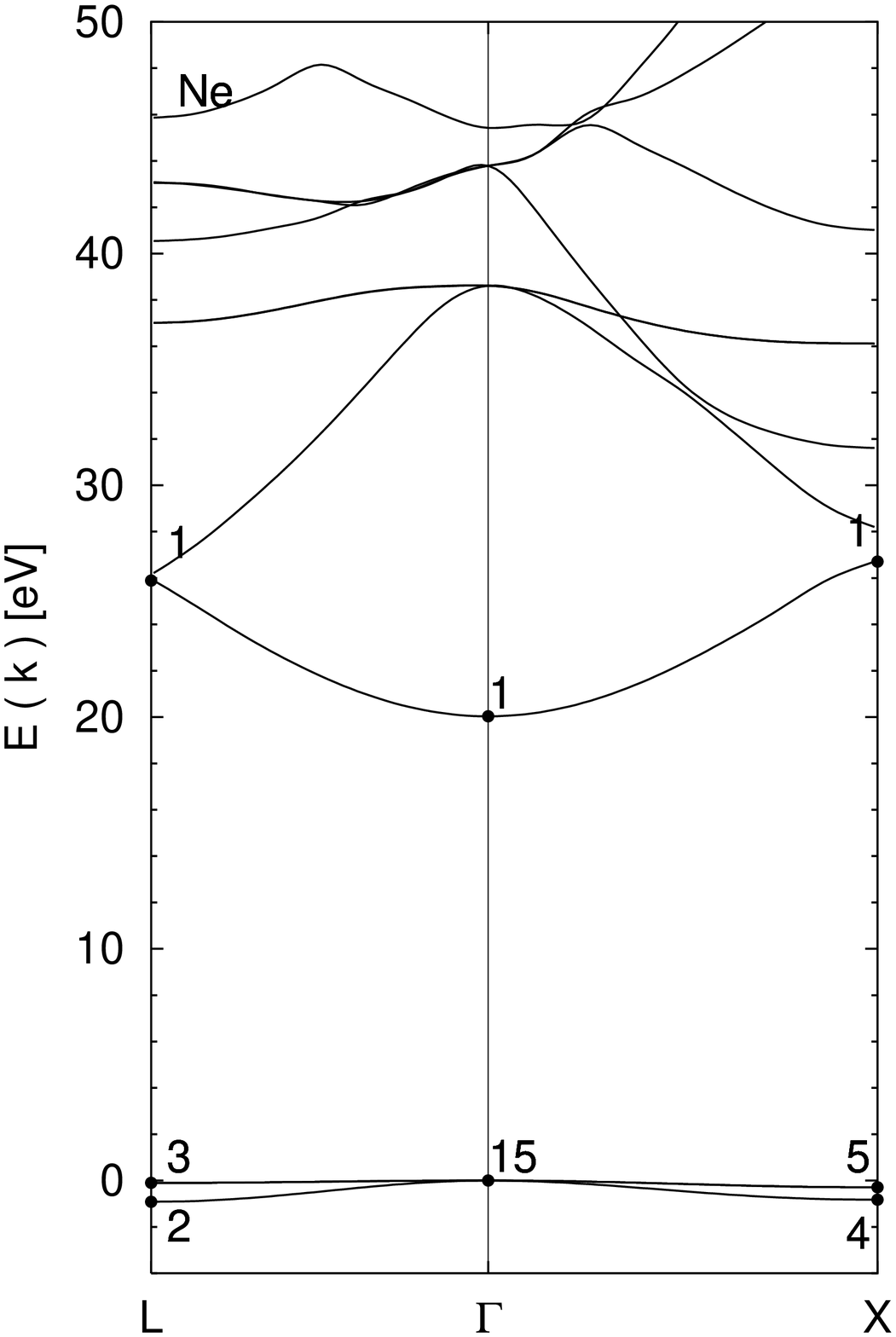}
\psfrag{Ar}[][c]{\colorbox{white}{Ar}}
\includegraphics[height=8 cm]{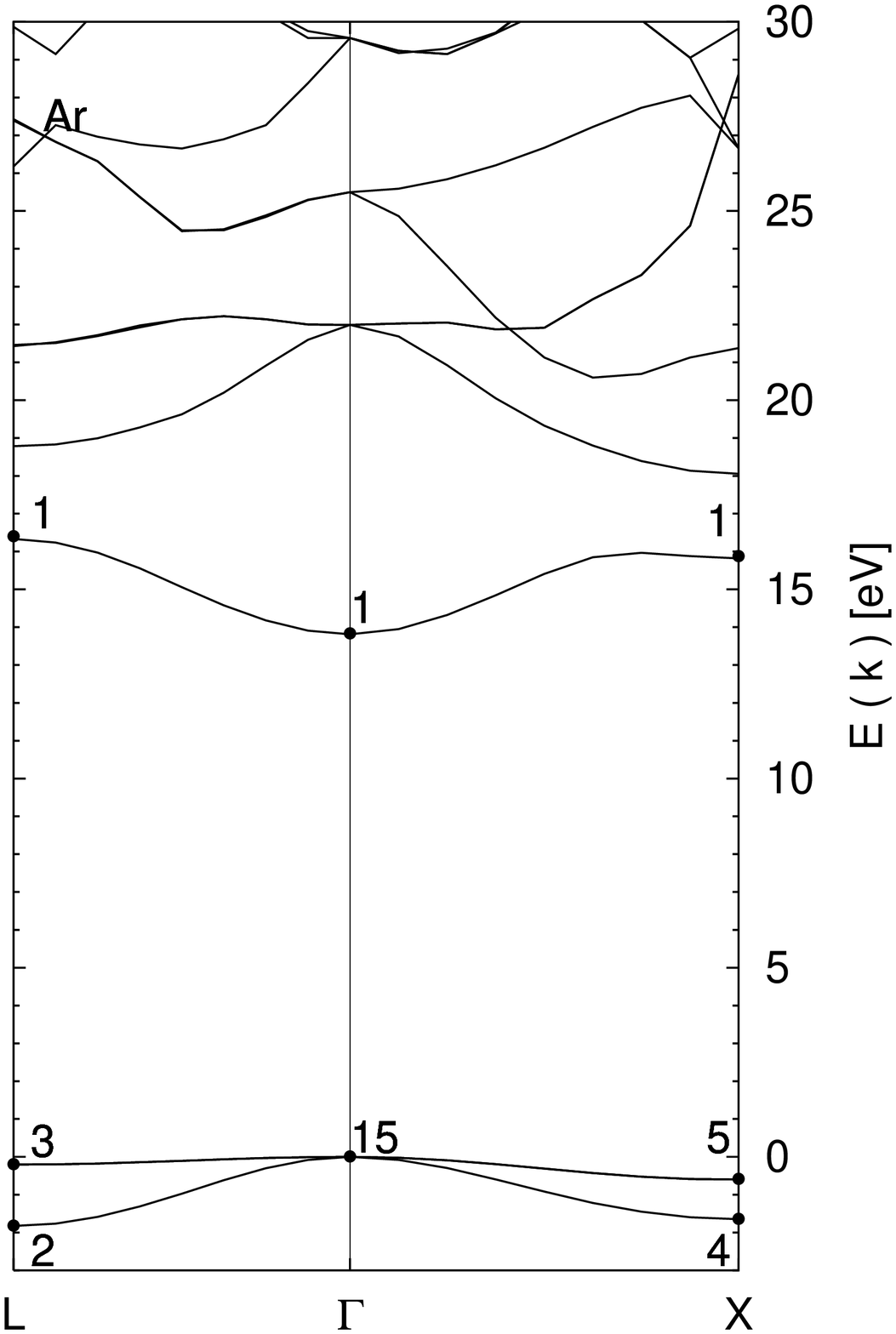}
\caption{\label{Bandstructure:F}$GW$ band structure for Ne (left panel) 
and Ar (right panel).}
\end{figure*}

$GW$ band structures along $\Delta$ and $\Sigma$ symmetry lines for Ne and Ar are shown in Fig.~\ref{Bandstructure:F}. Self-energy corrections to $GW$ band structures in both Ne and Ar are relatively independent of wavevector, leading to a \emph{scissor} type opening of the band gap on going from DFT to $GW$ energy bands. DFT bandstructures are not shown in Fig. \ref{Bandstructure:F} for clarity. Tables~\ref{BandsNe:T} and~\ref{BandsAr:T} give a direct comparison of all electron DFT energy eigenvalues and $GW$ quasiparticle energies at high symmetry points for Ne and Ar and include results from Bacalis \emph{et al.} \cite{Bacalis88} and experiment. When DFT and quasiparticle energies for Ne at X and L points are compared (Table \ref{BandsNe:T}), we find a widening of the valence bands by approximately 30\%. Our results for valence band energies and widths are in very good agreement with those reported earlier by Bacalis \emph{et al.} \cite{Bacalis88}. A similar pattern of valence band widening for $GW$ valence bands in Ne is found in Ar and our results are again in good agreement with those of Bacalis~\emph{et al.}.
\begin{table}
\caption{\label{BandsNe:T}Energy eigenvalues in eV at high-symmetry points for solid Ne. The reference energy is the valence band maximum energy. Results in the second and third columns were obtained using an all electron basis set and valence-core electron decoupling was done using the method outlined in Eq. \ref{Vxc_val:E}. Results in the fourth and fifth columns are from all electron PAW calculations \cite{Bacalis88}. The last column presents experimental values.}
\begin{ruledtabular}
\begin{tabular}{c d d d d d}
      &\multicolumn{2}{c}{This work} & %
       \multicolumn{2}{c}{PAW \footnote{Reference \onlinecite{Bacalis88}.} }\\
      & 
\multicolumn{1}{c}{DFT} & %
\multicolumn{1}{c}{$GW$}  & %
\multicolumn{1}{c}{DFT}                         & %
\multicolumn{1}{c}{$GW$}                      & %
\multicolumn{1}{c}{Exp.}  \\                  
\hline
$\Gamma _{15v}$                   &  0.0   & 0.0   &  0.0  &   0.0  &     0.0  \\
$\Gamma _{1c}$                    & 11.76  & 20.04 & 11.40 &  16.56 &    21.51\footnote{Reference \onlinecite{Runne95}.}\\
\multicolumn{5}{c}{}  \\
$X_{4v} ^{'}$                     & -0.61  & -0.82 & -0.67 &  -0.88 &    \\
$X_{5v} ^{'}$                     & -0.21  & -0.29 & -0.23 &  -0.30 &    \\
\multicolumn{5}{c}{}  \\
$L_{2v} ^{'}$                     &-0.69   & -0.91 & -0.75 &  -0.99 &   -1.3\footnote{Reference \onlinecite{Schwentner75}.}   \\
$L_{3v} ^{'}$                     &-0.07   & -0.10 & -0.07 &  -0.09 &      \\
\multicolumn{5}{c}{}  \\
$\Gamma _{25c} ^{'} - \Gamma _{1c}$& 17.85 & 18.57 &18.12   & 20.51    \\
$X_{1c} - \Gamma _{1c}$            &  6.78 & 6.66  & 6.82   &  8.12    \\
$L_{1c} - \Gamma _{1c}$            &  5.57 & 5.91  & 6.03   &  7.21    \\
\end{tabular} 
\end{ruledtabular}
\end{table}
The energy difference of the first and second conduction bands at the $\Gamma$ point ($\Gamma_{25c'}-\Gamma_{1c}$) in our $GW$ calculation is $\sim$ 2~eV smaller than in the PAW \cite{Bacalis88} calculation for Ne, while the value of 8.21~eV for Ar agrees well with the value of 8.44~eV obtained by Bacalis \etal. This energy difference is sensitive to completeness of Gaussian orbital basis sets (see Appendix~\ref{basisset:A}) as the $\Gamma_{25c'}$ state has significant amplitude in octahedral interstitial regions. Inclusion of interstitial functions in basis sets (Appendix~\ref{basisset:A}) and optimisation of the most diffuse functions reduced the $\Gamma_{25c'}-\Gamma_{1c}$ energy difference significantly, while basis sets with no interstitial functions result in a larger conduction band separation and fundamental gap.   
\begin{table}
\caption{\label{BandsAr:T} Energy eigenvalues in eV at high-symmetry points for solid Ar. The reference energy is the valence band maximum energy. Results in the second and third columns were obtained using an all electron basis set and valence-core electron decoupling was done using the method outlined in Eq. \ref{Vxc_val:E}. Results in the fourth and fifth columns are from all electron PAW calculations \cite{Bacalis88}. The last column presents experimental values.}
\begin{ruledtabular}
\begin{tabular}{c d d d d d}
      &\multicolumn{2}{c}{This work} & %
       \multicolumn{2}{c}{PAW \footnote{Reference \onlinecite{Bacalis88}.}} \\
      & 
\multicolumn{1}{c}{DFT} & %
\multicolumn{1}{c}{$GW$}  & %
\multicolumn{1}{c}{DFT}                         & %
\multicolumn{1}{c}{$GW$}                      & %
\multicolumn{1}{c}{Exp.}  \\                  
\hline 
$\Gamma _{15v}$   & 0.0 &  0.0   &  0.0   &  0.0    &  0.0   \\
$\Gamma _{1c}$    & 9.51&  13.82 &      8.09  & 11.96   &  14.15\footnote{Reference \onlinecite{Runne95}.}    \\
\multicolumn{5}{c}{}  \\
$X_{4v} ^{'}$     & -1.19 & -1.65  & -1.28  & -1.73   &     \\
$X_{5v} ^{'}$     & -0.42 & -0.50  & -0.46  & -0.63   &     \\
\multicolumn{5}{c}{}  \\
$L_{2v} ^{'}$     & -1.31 & -1.83 & -1.41  & -1.92   &  -1.7\footnote{Reference \onlinecite{Schwentner75}.}  \\
$L_{3v} ^{'}$     & -0.14  & -0.20  & -0.16  & -0.20   &    \\
\multicolumn{5}{c}{}  \\
$\Gamma _{25c} ^{'} - \Gamma _{1}$& 6.85 & 8.21  & 7.43     &   8.44    &\\
$X_{1c} - \Gamma _{1c}$           & 2.02 & 2.04  & 2.63     &   3.10    &\\
$L_{1c} - \Gamma _{1c}$           & 2.25 & 2.56  & 2.94     &   3.50    &\\
\end{tabular} 
\end{ruledtabular}
\end{table}

\section{Conclusions}

Band structures of solid Ne and Ar have been calculated using the $GW$ approximation. Calculations were performed using experimental lattice constants. Gaussian orbital basis sets were used throughout and core electrons were treated either explicitly with all electron basis sets or by pseudopotentials. Results of all electron and pseudopotential calculations are in good agreement, although the fundamental band gap predicted by all electron calculations is smaller than that in pseudopotential calculations by up to 0.2 eV. Positions of conduction band minima for Ne and Ar in $GW$ calculations are in good agreement with experimental electron affinities so that absolute positions of quasiparticle energy levels in Ne and Ar are reliably predicted in the $GW$ approximation. Fundamental band gaps for Ne and Ar are in good agreement with experimental gaps from photoemission and optical absorption data where shifts in the gap due to electron-hole attraction have been subtracted.

\begin{acknowledgments}

This work was supported by Enterprise-Ireland under Grant number SC/99/267 and by the Irish Higher Education Authority under the PRTLI-IITAC2 programme. CHP wishes to thank R. Dovesi and C. Roetti for hospitality during a visit to the Universit\`a di Torino, which was supported by a Royal Irish Academy Senior Visiting Fellowship  

\end{acknowledgments}

\appendix

\section{\label{basisset:A}Basis sets}

\begin{table}
\caption{\label{BasisSet:T}Basis sets used in this work. Exponents of s, p and d Cartesian Gaussian orbitals which were centered on the nuclear site ((0,0,0) labeled Nuc.) and at the octahedral interstitial site of the 
fcc lattice ((0.5,0.5,0.5) labelled Oct.) are tabulated in atomic units.  Basis sets for atomic cores in all electron calculations were conventional quantum 
chemistry core basis sets and are not given here.}
\begin{ruledtabular}
\begin{tabular}{c d d d d d d d}
Ne  &        &        &       &      &      &       &        \\
Nuc. sp  &  32.0  &  16.0  &  4.0  &  2.0 &  1.4 &  0.47 &  0.185 \\
Nuc.  d  &        &        &       &      &  1.6 &  0.8  &  0.2   \\
Oct. sp  &        &        &       &      &      &       &  0.3   \\
\\
\multicolumn{3}{l}{Ar - Basis Set 1} \\
Nuc. sp  &  32.0  &  16.0  &  4.0  &  2.0 &  1.4 &  0.47 &  0.15  \\
Nuc.  d  &        &        &       &      &  0.8 &  0.4  &  0.2   \\
Oct. sp  &        &        &       &      &      &  0.47 &  0.15  \\
Oct.  d  &        &        &       &      &      &       &  0.4   \\
\\
\multicolumn{3}{l}{Ar - Basis Set 2} \\
Nuc. sp  &  85.0  & 34.0   &  14.0 & 1.4   & 0.8  & 0.39 &  0.2  \\
Nuc.  d  &        &        &       &  1.05 & 0.79 & 0.39 &  0.1  \\                     
Oct. sp  &        &        &       &       &      & 0.61 &  0.31 \\
Oct   d  &        &        &       &       &      &      &  0.2  \\
\end{tabular} 
\end{ruledtabular}
\end{table}

The construction and use of an appropriate basis set constitutes a critical factor in \ab calculations and is particularly important within a Gaussian orbital framework. Apart from minimizing the total energy, which is necessary for a good quality basis set, one has to ensure that the basis set contains a sufficient number of basis functions to generate the virtual space. The need for a large number of conduction bands for a well converged self-energy has been emphasized again recently \cite{Tiago04}. The number of conduction bands can be increased by including $f$ and $g$ type functions into the basis set, but these are not yet available in the CRYSTAL code. Alternatively, extra sets of orbitals were added at interstitial sites of the crystal. This improves the flexibility of the basis set through the unit cell and attempts to reproduce the highly nodal structure of free-electron conduction band states. 

Two techniques were used for constructing basis sets: Firstly, starting from two decay constants, 0.15 and 2.0, geometrical expansion was used to generate more localized orbitals, interstitial functions were added and the most diffuse functions were adjusted to minimize the total energy. The second approach used valence exponents from conventional, contracted quantum chemistry basis sets. Several Gaussian functions are combined into a single basis function in a contracted basis function by fixing their weights. Here the same exponents as used in contracted basis functions were used, but relative weights of different exponents were determined during the self-consistent field DFT calculation. The basis set used for PP Ne and Ar (Basis set 1) and all electron Ne calculations was of the first type while all electron Ar calculations were performed using a basis set of the second type (Basis set 2).

\FloatBarrier

\bibliography{paper2}

\begin{thebibliography}{37}
\expandafter\ifx\csname natexlab\endcsname\relax\def\natexlab#1{#1}\fi
\expandafter\ifx\csname bibnamefont\endcsname\relax
  \def\bibnamefont#1{#1}\fi
\expandafter\ifx\csname bibfnamefont\endcsname\relax
  \def\bibfnamefont#1{#1}\fi
\expandafter\ifx\csname citenamefont\endcsname\relax
  \def\citenamefont#1{#1}\fi
\expandafter\ifx\csname url\endcsname\relax
  \def\url#1{\texttt{#1}}\fi
\expandafter\ifx\csname urlprefix\endcsname\relax\def\urlprefix{URL }\fi
\providecommand{\bibinfo}[2]{#2}
\providecommand{\eprint}[2][]{\url{#2}}

\bibitem[{\citenamefont{Lipari and Fowler}(1970)}]{Lipari70}
\bibinfo{author}{\bibfnamefont{N.~O.} \bibnamefont{Lipari}} \bibnamefont{and}
  \bibinfo{author}{\bibfnamefont{W.~B.} \bibnamefont{Fowler}},
  \bibinfo{journal}{Phys. Rev. B} \textbf{\bibinfo{volume}{2}},
  \bibinfo{pages}{3354} (\bibinfo{year}{1970}).

\bibitem[{\citenamefont{Bacalis et~al.}(1988)\citenamefont{Bacalis,
  Papaconstantopoulos, and Pickett}}]{Bacalis88}
\bibinfo{author}{\bibfnamefont{N.~C.} \bibnamefont{Bacalis}},
  \bibinfo{author}{\bibfnamefont{D.~A.} \bibnamefont{Papaconstantopoulos}},
  \bibnamefont{and} \bibinfo{author}{\bibfnamefont{W.~E.}
  \bibnamefont{Pickett}}, \bibinfo{journal}{Phys. Rev. B}
  \textbf{\bibinfo{volume}{38}}, \bibinfo{pages}{6218} (\bibinfo{year}{1988}).

\bibitem[{\citenamefont{Magyar et~al.}(2004)\citenamefont{Magyar, Fleszar, and
  Gross}}]{Magyar04}
\bibinfo{author}{\bibfnamefont{R.}~\bibnamefont{Magyar}},
  \bibinfo{author}{\bibfnamefont{A.}~\bibnamefont{Fleszar}}, \bibnamefont{and}
  \bibinfo{author}{\bibfnamefont{E.}~\bibnamefont{Gross}},
  \bibinfo{journal}{Phys. Rev. B} \textbf{\bibinfo{volume}{69}},
  \bibinfo{pages}{045111} (\bibinfo{year}{2004}).

\bibitem[{\citenamefont{Dagens and Perrot}(1972)}]{Dagens72}
\bibinfo{author}{\bibfnamefont{L.}~\bibnamefont{Dagens}} \bibnamefont{and}
  \bibinfo{author}{\bibfnamefont{F.}~\bibnamefont{Perrot}},
  \bibinfo{journal}{Phys. Rev. B} \textbf{\bibinfo{volume}{5}},
  \bibinfo{pages}{641} (\bibinfo{year}{1972}).

\bibitem[{\citenamefont{Kunz and Mickish}(1973)}]{Kunz73}
\bibinfo{author}{\bibfnamefont{A.~B.} \bibnamefont{Kunz}} \bibnamefont{and}
  \bibinfo{author}{\bibfnamefont{D.~J.} \bibnamefont{Mickish}},
  \bibinfo{journal}{Phys. Rev. B} \textbf{\bibinfo{volume}{8}},
  \bibinfo{pages}{779} (\bibinfo{year}{1973}).

\bibitem[{\citenamefont{Baroni et~al.}(1981)\citenamefont{Baroni, Grosso, and
  Parravicini}}]{Baroni81}
\bibinfo{author}{\bibfnamefont{S.}~\bibnamefont{Baroni}},
  \bibinfo{author}{\bibfnamefont{G.}~\bibnamefont{Grosso}}, \bibnamefont{and}
  \bibinfo{author}{\bibfnamefont{G.~P.} \bibnamefont{Parravicini}},
  \bibinfo{journal}{Phys. Rev. B} \textbf{\bibinfo{volume}{23}},
  \bibinfo{pages}{6441} (\bibinfo{year}{1981}).

\bibitem[{\citenamefont{Baroni et~al.}(1984)\citenamefont{Baroni, Grosso, and
  Parravicini}}]{Baroni84}
\bibinfo{author}{\bibfnamefont{S.}~\bibnamefont{Baroni}},
  \bibinfo{author}{\bibfnamefont{G.}~\bibnamefont{Grosso}}, \bibnamefont{and}
  \bibinfo{author}{\bibfnamefont{G.~P.} \bibnamefont{Parravicini}},
  \bibinfo{journal}{Phys. Rev. B} \textbf{\bibinfo{volume}{29}},
  \bibinfo{pages}{2891} (\bibinfo{year}{1984}).

\bibitem[{\citenamefont{Heaton et~al.}(1983)\citenamefont{Heaton, Harrison, and
  Lin}}]{Heaton83}
\bibinfo{author}{\bibfnamefont{R.~A.} \bibnamefont{Heaton}},
  \bibinfo{author}{\bibfnamefont{J.}~\bibnamefont{Harrison}}, \bibnamefont{and}
  \bibinfo{author}{\bibfnamefont{C.}~\bibnamefont{Lin}},
  \bibinfo{journal}{Phys. Rev. B} \textbf{\bibinfo{volume}{28}},
  \bibinfo{pages}{5992} (\bibinfo{year}{1983}).

\bibitem[{\citenamefont{Rohlfing and Louie}(2000)}]{Rohlfing00}
\bibinfo{author}{\bibfnamefont{M.}~\bibnamefont{Rohlfing}} \bibnamefont{and}
  \bibinfo{author}{\bibfnamefont{S.~G.} \bibnamefont{Louie}},
  \bibinfo{journal}{Phys. Rev. B} \textbf{\bibinfo{volume}{62}},
  \bibinfo{pages}{4927} (\bibinfo{year}{2000}).

\bibitem[{\citenamefont{Hedin and Lundqvist}(1969)}]{Hedin69}
\bibinfo{author}{\bibfnamefont{L.}~\bibnamefont{Hedin}} \bibnamefont{and}
  \bibinfo{author}{\bibfnamefont{L.}~\bibnamefont{Lundqvist}},
  \emph{\bibinfo{title}{Solid State Physics}}, vol.~\bibinfo{volume}{23}
  (\bibinfo{publisher}{Academic Press}, \bibinfo{address}{New York},
  \bibinfo{year}{1969}).

\bibitem[{\citenamefont{Hybertsen and Louie}(1986)}]{Hybertsen86}
\bibinfo{author}{\bibfnamefont{M.~S.} \bibnamefont{Hybertsen}}
  \bibnamefont{and} \bibinfo{author}{\bibfnamefont{S.~G.} \bibnamefont{Louie}},
  \bibinfo{journal}{Phys. Rev. B} \textbf{\bibinfo{volume}{34}},
  \bibinfo{pages}{5390} (\bibinfo{year}{1986}).

\bibitem[{\citenamefont{Galami{\'c}-Mulaomerovi{\'c} and
  Patterson}()}]{Galamic04a}
\bibinfo{author}{\bibfnamefont{S.}~\bibnamefont{Galami{\'c}-Mulaomerovi{\'c}}}
  \bibnamefont{and} \bibinfo{author}{\bibfnamefont{C.~H.}
  \bibnamefont{Patterson}}, \eprint{in preparation}.

\bibitem[{\citenamefont{Galami{\'c}-Mulaomerovi{\'c}
  et~al.}()\citenamefont{Galami{\'c}-Mulaomerovi{\'c}, Hogan, and
  Patterson}}]{Galamic04}
\bibinfo{author}{\bibfnamefont{S.}~\bibnamefont{Galami{\'c}-Mulaomerovi{\'c}}},
  \bibinfo{author}{\bibfnamefont{C.~D.} \bibnamefont{Hogan}}, \bibnamefont{and}
  \bibinfo{author}{\bibfnamefont{C.~H.} \bibnamefont{Patterson}}, \eprint{in
  preparation}.

\bibitem[{\citenamefont{Saunders et~al.}(2003)\citenamefont{Saunders, Dovesi,
  Roetti, Caus\'a, Orlando, Zicovich-Wilson, Harrison, Doll, Civalleri, Bush
  et~al.}}]{Crystal03}
\bibinfo{author}{\bibfnamefont{V.~R.} \bibnamefont{Saunders}},
  \bibinfo{author}{\bibfnamefont{R.}~\bibnamefont{Dovesi}},
  \bibinfo{author}{\bibfnamefont{C.}~\bibnamefont{Roetti}},
  \bibinfo{author}{\bibfnamefont{M.}~\bibnamefont{Caus\'a}},
  \bibinfo{author}{\bibfnamefont{R.}~\bibnamefont{Orlando}},
  \bibinfo{author}{\bibfnamefont{C.~M.} \bibnamefont{Zicovich-Wilson}},
  \bibinfo{author}{\bibfnamefont{N.~M.} \bibnamefont{Harrison}},
  \bibinfo{author}{\bibfnamefont{K.}~\bibnamefont{Doll}},
  \bibinfo{author}{\bibfnamefont{B.}~\bibnamefont{Civalleri}},
  \bibinfo{author}{\bibfnamefont{I.}~\bibnamefont{Bush}}, \bibnamefont{et~al.},
  \emph{\bibinfo{title}{CRYSTAL03 User's Manual}}, \bibinfo{address}{Torino}
  (\bibinfo{year}{2003}).

\bibitem[{\citenamefont{Runne and Zimmerer}(1995)}]{Runne95}
\bibinfo{author}{\bibfnamefont{M.}~\bibnamefont{Runne}} \bibnamefont{and}
  \bibinfo{author}{\bibfnamefont{G.}~\bibnamefont{Zimmerer}},
  \bibinfo{journal}{Nucl. Instr. Meth. B} \textbf{\bibinfo{volume}{101}},
  \bibinfo{pages}{156} (\bibinfo{year}{1995}).

\bibitem[{\citenamefont{Schwentner et~al.}(1973)\citenamefont{Schwentner,
  Skibowski, and Steinmann}}]{Schwentner73}
\bibinfo{author}{\bibfnamefont{N.}~\bibnamefont{Schwentner}},
  \bibinfo{author}{\bibfnamefont{M.}~\bibnamefont{Skibowski}},
  \bibnamefont{and}
  \bibinfo{author}{\bibfnamefont{W.}~\bibnamefont{Steinmann}},
  \bibinfo{journal}{Phys. Rev. B} \textbf{\bibinfo{volume}{8}},
  \bibinfo{pages}{2965} (\bibinfo{year}{1973}).

\bibitem[{\citenamefont{Schwentner et~al.}(1975)\citenamefont{Schwentner,
  Himpsel, Savle, Skibowski, Steinmann, and Koch}}]{Schwentner75}
\bibinfo{author}{\bibfnamefont{N.}~\bibnamefont{Schwentner}},
  \bibinfo{author}{\bibfnamefont{F.-J.} \bibnamefont{Himpsel}},
  \bibinfo{author}{\bibfnamefont{V.}~\bibnamefont{Savle}},
  \bibinfo{author}{\bibfnamefont{M.}~\bibnamefont{Skibowski}},
  \bibinfo{author}{\bibfnamefont{W.}~\bibnamefont{Steinmann}},
  \bibnamefont{and} \bibinfo{author}{\bibfnamefont{E.}~\bibnamefont{Koch}},
  \bibinfo{journal}{Phys. Rev. Lett.} \textbf{\bibinfo{volume}{34}},
  \bibinfo{pages}{528} (\bibinfo{year}{1975}).

\bibitem[{\citenamefont{Hamada et~al.}(1990)\citenamefont{Hamada, Hwang, and
  Freeman}}]{Hamada90}
\bibinfo{author}{\bibfnamefont{N.}~\bibnamefont{Hamada}},
  \bibinfo{author}{\bibfnamefont{M.}~\bibnamefont{Hwang}}, \bibnamefont{and}
  \bibinfo{author}{\bibfnamefont{A.}~\bibnamefont{Freeman}},
  \bibinfo{journal}{Phys. Rev. B} \textbf{\bibinfo{volume}{41}},
  \bibinfo{pages}{3620} (\bibinfo{year}{1990}).

\bibitem[{\citenamefont{Arnaud and Alouani}(2000)}]{Arnaud00}
\bibinfo{author}{\bibfnamefont{B.}~\bibnamefont{Arnaud}} \bibnamefont{and}
  \bibinfo{author}{\bibfnamefont{M.}~\bibnamefont{Alouani}},
  \bibinfo{journal}{Phys. Rev. B} \textbf{\bibinfo{volume}{62}},
  \bibinfo{pages}{4464} (\bibinfo{year}{2000}).

\bibitem[{\citenamefont{Kotani and van Schilfgaarde}(2002)}]{Kotani02}
\bibinfo{author}{\bibfnamefont{T.}~\bibnamefont{Kotani}} \bibnamefont{and}
  \bibinfo{author}{\bibfnamefont{M.}~\bibnamefont{van Schilfgaarde}},
  \bibinfo{journal}{Solid State Commun.} \textbf{\bibinfo{volume}{121}},
  \bibinfo{pages}{461} (\bibinfo{year}{2002}).

\bibitem[{\citenamefont{Ku and Eguiluz}(2003)}]{Ku02}
\bibinfo{author}{\bibfnamefont{W.}~\bibnamefont{Ku}} \bibnamefont{and}
  \bibinfo{author}{\bibfnamefont{A.}~\bibnamefont{Eguiluz}},
  \bibinfo{journal}{Phys. Rev. Lett.} \textbf{\bibinfo{volume}{89}},
  \bibinfo{pages}{126401} (\bibinfo{year}{2003}).

\bibitem[{\citenamefont{Tiago et~al.}(2004)\citenamefont{Tiago, Ismail-Beigi,
  and Louie}}]{Tiago04}
\bibinfo{author}{\bibfnamefont{M.~L.} \bibnamefont{Tiago}},
  \bibinfo{author}{\bibfnamefont{S.}~\bibnamefont{Ismail-Beigi}},
  \bibnamefont{and} \bibinfo{author}{\bibfnamefont{S.~G.} \bibnamefont{Louie}},
  \bibinfo{journal}{Phys. Rev. B} \textbf{\bibinfo{volume}{69}},
  \bibinfo{pages}{125212} (\bibinfo{year}{2004}).

\bibitem[{\citenamefont{Leb{\`e}gue et~al.}(2003)\citenamefont{Leb{\`e}gue,
  Arnaud, Alouani, and Bloechl}}]{Lebegue03}
\bibinfo{author}{\bibfnamefont{S.}~\bibnamefont{Leb{\`e}gue}},
  \bibinfo{author}{\bibfnamefont{B.}~\bibnamefont{Arnaud}},
  \bibinfo{author}{\bibfnamefont{M.}~\bibnamefont{Alouani}}, \bibnamefont{and}
  \bibinfo{author}{\bibfnamefont{P.~E.} \bibnamefont{Bloechl}},
  \bibinfo{journal}{Phys. Rev. B} \textbf{\bibinfo{volume}{67}},
  \bibinfo{pages}{155208} (\bibinfo{year}{2003}).

\bibitem[{\citenamefont{Hedin}(1965)}]{Hedin65}
\bibinfo{author}{\bibfnamefont{L.}~\bibnamefont{Hedin}},
  \bibinfo{journal}{Phys. Rev.} \textbf{\bibinfo{volume}{139}},
  \bibinfo{pages}{A796} (\bibinfo{year}{1965}).

\bibitem[{\citenamefont{Godby et~al.}(1988)\citenamefont{Godby, Schl{\"u}ter,
  and Sham}}]{Godby88}
\bibinfo{author}{\bibfnamefont{R.~W.} \bibnamefont{Godby}},
  \bibinfo{author}{\bibfnamefont{M.}~\bibnamefont{Schl{\"u}ter}},
  \bibnamefont{and} \bibinfo{author}{\bibfnamefont{L.~J.} \bibnamefont{Sham}},
  \bibinfo{journal}{Phys. Rev. B} \textbf{\bibinfo{volume}{37}},
  \bibinfo{pages}{10159} (\bibinfo{year}{1988}).

\bibitem[{\citenamefont{von~der Linden and Horsch}(1988)}]{vonderLinden88}
\bibinfo{author}{\bibfnamefont{W.}~\bibnamefont{von~der Linden}}
  \bibnamefont{and} \bibinfo{author}{\bibfnamefont{P.}~\bibnamefont{Horsch}},
  \bibinfo{journal}{Phys. Rev. B} \textbf{\bibinfo{volume}{37}},
  \bibinfo{pages}{8351} (\bibinfo{year}{1988}).

\bibitem[{\citenamefont{Baldereschi and Tosatti}(1979)}]{Baldereschi79}
\bibinfo{author}{\bibfnamefont{A.}~\bibnamefont{Baldereschi}} \bibnamefont{and}
  \bibinfo{author}{\bibfnamefont{E.}~\bibnamefont{Tosatti}},
  \bibinfo{journal}{Sol. St. Commun.} \textbf{\bibinfo{volume}{29}},
  \bibinfo{pages}{131} (\bibinfo{year}{1979}).

\bibitem[{\citenamefont{Galami{\'c}-Mulaomerovi{\'c}
  et~al.}(2001)\citenamefont{Galami{\'c}-Mulaomerovi{\'c}, Hogan, and
  Patterson}}]{Galamic01}
\bibinfo{author}{\bibfnamefont{S.}~\bibnamefont{Galami{\'c}-Mulaomerovi{\'c}}},
  \bibinfo{author}{\bibfnamefont{C.~D.} \bibnamefont{Hogan}}, \bibnamefont{and}
  \bibinfo{author}{\bibfnamefont{C.~H.} \bibnamefont{Patterson}},
  \bibinfo{journal}{Phys. Stat. Sol. (a)} \textbf{\bibinfo{volume}{188}},
  \bibinfo{pages}{1291} (\bibinfo{year}{2001}).

\bibitem[{\citenamefont{Johnson}(1974)}]{Johnson74}
\bibinfo{author}{\bibfnamefont{D.~L.} \bibnamefont{Johnson}},
  \bibinfo{journal}{Phys. Rev. B} \textbf{\bibinfo{volume}{9}},
  \bibinfo{pages}{4475} (\bibinfo{year}{1974}).

\bibitem[{\citenamefont{Hott}(1991)}]{Hott91}
\bibinfo{author}{\bibfnamefont{R.}~\bibnamefont{Hott}}, \bibinfo{journal}{Phys.
  Rev. B} \textbf{\bibinfo{volume}{44}}, \bibinfo{pages}{1057}
  (\bibinfo{year}{1991}).

\bibitem[{\citenamefont{Hohenberg and Kohn}(1964)}]{KohnPR:136}
\bibinfo{author}{\bibfnamefont{P.}~\bibnamefont{Hohenberg}} \bibnamefont{and}
  \bibinfo{author}{\bibfnamefont{W.}~\bibnamefont{Kohn}}, \bibinfo{journal}{PR}
  \textbf{\bibinfo{volume}{136}}, \bibinfo{pages}{B864} (\bibinfo{year}{1964}).

\bibitem[{\citenamefont{Perdew and Wang}(1992)}]{Perdew92}
\bibinfo{author}{\bibfnamefont{J.}~\bibnamefont{Perdew}} \bibnamefont{and}
  \bibinfo{author}{\bibfnamefont{Y.}~\bibnamefont{Wang}},
  \bibinfo{journal}{Phys. Rev. B} \textbf{\bibinfo{volume}{45}},
  \bibinfo{pages}{13244} (\bibinfo{year}{1992}).

\bibitem[{\citenamefont{Barthelat et~al.}(1977)\citenamefont{Barthelat, Durand,
  and Serafini}}]{Barthelat77}
\bibinfo{author}{\bibfnamefont{J.}~\bibnamefont{Barthelat}},
  \bibinfo{author}{\bibfnamefont{P.}~\bibnamefont{Durand}}, \bibnamefont{and}
  \bibinfo{author}{\bibfnamefont{A.}~\bibnamefont{Serafini}},
  \bibinfo{journal}{Mol. Phys.} \textbf{\bibinfo{volume}{33}},
  \bibinfo{pages}{159} (\bibinfo{year}{1977}).

\bibitem[{\citenamefont{Sonntag}(1977)}]{Sonntag77}
\bibinfo{author}{\bibfnamefont{B.}~\bibnamefont{Sonntag}},
  \emph{\bibinfo{title}{Rare Gas Solids}}, vol.~\bibinfo{volume}{2}
  (\bibinfo{publisher}{Eds. M.L. Klein and J. A. Venables, Academic Press},
  \bibinfo{address}{London}, \bibinfo{year}{1977}).

\bibitem[{\citenamefont{Monkhorst and Pack}(1976)}]{Monkhorst76}
\bibinfo{author}{\bibfnamefont{H.~J.} \bibnamefont{Monkhorst}}
  \bibnamefont{and} \bibinfo{author}{\bibfnamefont{J.~D.} \bibnamefont{Pack}},
  \bibinfo{journal}{Phys. Rev. B} \textbf{\bibinfo{volume}{13}},
  \bibinfo{pages}{5188} (\bibinfo{year}{1976}).

\bibitem[{\citenamefont{Gygi and Baldereschi}(1986)}]{Gygi86}
\bibinfo{author}{\bibfnamefont{F.}~\bibnamefont{Gygi}} \bibnamefont{and}
  \bibinfo{author}{\bibfnamefont{A.}~\bibnamefont{Baldereschi}},
  \bibinfo{journal}{Phys. Rev. B} \textbf{\bibinfo{volume}{34}},
  \bibinfo{pages}{4405} (\bibinfo{year}{1986}).

\bibitem[{\citenamefont{Shirley et~al.}(1997)\citenamefont{Shirley, Zhu, and
  Louie}}]{Shirley97}
\bibinfo{author}{\bibfnamefont{E.~L.} \bibnamefont{Shirley}},
  \bibinfo{author}{\bibfnamefont{X.}~\bibnamefont{Zhu}}, \bibnamefont{and}
  \bibinfo{author}{\bibfnamefont{S.~G.} \bibnamefont{Louie}},
  \bibinfo{journal}{Phys. Rev. B} \textbf{\bibinfo{volume}{56}},
  \bibinfo{pages}{6648} (\bibinfo{year}{1997}).

\end{thebibliography}

\end{document}